\documentclass{ws-ijmpd}
\usepackage[super,compress]{cite}
\usepackage[colorlinks,citecolor=blue, linkcolor=blue]{hyperref}
\begin{document}

\markboth{W. Zhao  et al.}
{Ultralight scalar dark matter detection with ZAIGA}

%
\catchline{}{}{}{}{}
%

\title{Ultralight scalar dark matter detection with ZAIGA}
\author{Wei Zhao\textsuperscript{1,2}, Xitong Mei\textsuperscript{1,2}, Dongfeng Gao\textsuperscript{1,$^\ast$}, Jin Wang\textsuperscript{1,$\dagger$}, Mingsheng Zhan\textsuperscript{1,$\ddagger$}}
\address{\textsuperscript{1}State Key Laboratory of Magnetic Resonance and Atomic and Molecular Physics,
\\Wuhan Institute of Physics and Mathematics, APM,\\
	Chinese Academy of Sciences, \\Wuhan 430071, China
	\\ \textsuperscript{2}School of Physical Sciences, \\University of Chinese Academy of Sciences, \\Beijing 100049, China
	\\$^\ast$dfgao@wipm.ac.cn\\$^\dagger$wangjin@wipm.ac.cn\\
$^\ddagger$mszhan@wipm.ac.cn
	}

\maketitle

\begin{history}
\received{Day Month Year}
\revised{Day Month Year}
\end{history}

\begin{abstract}
   ZAIGA is a proposed underground long-baseline atom interferometer (AI) facility, aiming for experimental research on gravitation and related problems. In this paper, we study the possibility of detecting the ultralight scalar dark matter (DM) with ZAIGA. According to a popular scalar DM model, the DM field contains a background oscillation term and a local exponential fluctuation term. In order to calculate the proposed constraints on DM coupling parameters, we need to first compute the DM signals in ZAIGA. For the case of two AIs vertically separated by 300 meters, the DM-induced differential phase consists of three contributions, coming from the DM-induced changes in atomic internal energy levels, atomic masses and the gravitational acceleration. For the case of two AIs horizontally separated by several kilometers, the signal comes from the DM-induced changes in atomic internal energy levels. With the current and future technical parameters of ZAIGA, we then obtain the proposed constraints on five DM coupling parameters. It turns out that our proposed constraints could be several orders of magnitude better than the ones set by the MICROSCOPE space mission.
\end{abstract}


\ccode{PACS numbers:}


\section{Introduction}	
Dark matter (DM) is one of the most challenging problems of modern physics and cosmology. A variety of astrophysical and cosmological observations indicate the existence of DM \cite{1,2,3}. Current data suggest that 80\% of all matters in the universe is DM \cite{cosmological}. Up to now, we only observe the gravitational effect of DM, while its other properties are still unknown.
Weakly interacting massive particles (WIMPs) are the major DM candidate but no signals have been found \cite{RN405,RN403,RN404}. Recently, the ultralight DM candidate attracts a lot of attentions. There are various proposals to search for the ultralight DM with precision tools, such as AIs \cite{RN17}, atomic clocks \cite{Derevianko14,RN2,Roberts20,Beloy21Nature:ACDM}, accelerometers \cite{RN45,Manley21PRL:OptomechanicalAccelerometerVectorDM,Carney21}, optical cavities \cite{Obata18,RN129,Savalle21,Kennedy20} and laser interferometers \cite{Stadnik15:LaserInterferometry,Stadnik16,Grote19PRR:GWDM,Pierce18PRLDPGW,Nagano19,Guo19:DPLIGO,Morisaki19,Morisaki21PRD:GWvectorDM,Michimura20PRD:GWvectorDM}.

AIs rely on coherently manipulating atomic matter waves. Details could be found in the paper \cite{RN78}. AIs have been widely applied to test the weak equivalence principle with accuracy of $10^{-12}$-level \cite{RN190} and measure the fine structure constant  with $\alpha^{-1}$= 137.035999206(11) \cite{RN194}. More applications could be found in the paper \cite{RN286}. Recently, several long-baseline atomic sensor schemes have been put forward, such as AION \cite{RN255}, MAGIS-100 \cite{RN247}, MIGA \cite{RN187}, ELGAR \cite{RN256} and ZAIGA \cite{RN106}. One important goal of these schemes is to search for the ultralight DM.

According to the popular ultralight scalar DM model \cite{RN18,RN16}, DM may interact with standard model matters and change the mass of fermions, the electromagnetic fine structure constant as well as the QCD energy scale. Consequently, the masses of atoms and the Earth will be modified by the DM. These modifications will go into the phase shift of AIs. We have given a complete result for the DM-induced phase shift for a single AI in the paper \cite{RN110}. In this paper, we will first compute the DM signals in ZAIGA, namely for the case of two vertically separated AIs, and  the case of two horizontally separated AIs. After that, we compute the proposed constraints on five DM coupling parameters and compare with the MICROSCOPE space mission.
  
The paper is organized as follows. We first introduce the popular ultralight scalar DM model and give the DM-induced phase shift for a single AI in Sec. \ref{the scalar dark matter model}. In Sec. \ref{The ultralight scalar DM detection with ZAIGA}, we introduce the ZAIGA proposal and its technical parameters for detecting the ultralight scalar DM. The DM-induced differential phases for two separated AIs are also  calculated. Then we discuss the possible constraints on five coupling parameters by ZAIGA in Sec. \ref{Constraints on the coupling parameters}. Finally, conclusion and discussion are made in Sec. \ref{Conclusion and discussion}.

\section{The ultralight scalar DM model \label{the scalar dark matter model}}
In this section, we will briefly introduce  the ultralight scalar DM model    \cite{RN18,RN16} and the result of DM-induced phase shift for a single AI calculated in our previous paper \cite{RN110}.

\subsection{The model}
For the linear coupling DM model, the action and the interaction Lagrangian density at the microscopic level are the following
\begin{align}
S &=\int d^{4}x \dfrac{\sqrt{-g}}{2\kappa}\Bigg[R-2g^{\mu\nu}\partial_{\mu}\varphi \partial_{\nu}\varphi-V(\varphi)\Bigg]
\notag \\
&+\int d^{4}x\sqrt{-g}\Bigg[\mathcal{L}_{SM}(g_{\mu\nu},\psi_{i})+\mathcal{L}_{int}(g_{\mu\nu},\varphi,\psi_{i})\Bigg] \, ,\label{1}
\end{align}
and 
\begin{align}
\mathcal{L}_{int}&=\varphi \Bigg[  \frac{d_{e}}{4 e^2}F_{\mu\nu}F^{\mu\nu}
-\frac{d_{g}\beta_{3}}{2g_{3}}F^A_{\mu\nu}F^{A\mu\nu}  
  - \sum_{i=e,u,d} (d_{m_{i}}+\gamma_{m_{i}}d_{g})m_{i}\bar{\psi}_{i}\psi_{i} \Bigg] \, ,
\end{align}
where $\kappa=\frac{8\pi G}{c^{4}}$, $g_{\mu\nu}$ is the spacetime metric, and $\varphi$ denotes the dimensionless scalar DM field. $\beta_{3}$ is the beta function for $g_{3}$, where $g_3$ is the QCD gauge coupling. $\gamma_{m_{i}}$ is the anomalous dimension from the energy running of the quark masses, $d_{e}$, $d_{g}$ and $d_{m_{i}}$ are the coupling parameters  between DM and the electromagnetic field, the gluonic field as well as the masses of the electron
and quarks. For convenience, we write the symmetrical and  antisymmetric form for the coupling parameters of up and down quarks as
\begin{align}
	d_{\hat{m}} =\frac{m_ud_{m_u}+m_dd_{m_d}}{m_u+m_d} \, ,  \ 
	d_{\delta m} =\frac{m_dd_{m_d}-m_ud_{m_u}}{m_d-m_u}\, . \label{5}
\end{align}
To describe the behavior of the DM field near the Earth, the following phenomenological action is derived from the microscopic action Eq. (\ref{1})
\begin{equation}
S =\int d^{4}x \dfrac{\sqrt{-g}}{2\kappa}\Bigg[R-2g^{\mu\nu}\partial_{\mu}\varphi \partial_{\nu}\varphi-V(\varphi)\Bigg]-c^2 \int\rho_{E}(\varphi)\, \sqrt{-g}\, d^4 x \, ,  \label{earthaction}
\end{equation}
where $V({\varphi})=2\frac{c^2m_{\varphi}^2}{\hbar^2}\varphi^{2}$ is the quadratic scalar potential,  $\rho_{E}=\frac{3M_{E}(\varphi)}{4\pi R_{E}^{3}}$ is the average density of the Earth. We can do the Taylor expansion for the Earth's mass  
\begin{equation}
M_{E}(\varphi)=M_E \Bigg[1+\alpha_E\varphi+\tilde{\alpha_{E}}\varphi^2 + \mathcal{O}(\varphi^3)\Bigg]\, .
\end{equation} 
Here 
$
\alpha_{E}=0.92d_g+0.08d_{\hat{m}}+2.35\times 10^{-5}d_{\delta m}
+2.71\times 10^{-4}d_{m_e}+1.71\times 10^{-3}d_e 
$
is the DM charge for the Earth and $\tilde{\alpha_{E}}\backsimeq d_{g}^{\, 2}$ \cite{RN110}.

From Eq. (\ref{earthaction}), one can get the solution of $\varphi$ near the Earth
\begin{equation}
\varphi=\varphi_{0} \cos(k_{\varphi}\, r-\omega t+\delta)-\alpha_{E}I(\frac{R_E}{\lambda_{\text{eff}}})\frac{GM_{E}}{c^2}\frac{ e^{-\frac{r}{\lambda_{\text{eff}}}}}{r}\, ,\label{solution of DM} 
\end{equation} 
where $\varphi_{0}=\frac{7.2\times 10^{-31} \rm eV}{m_{\varphi}}$ is the amplitude, $k_{\varphi}=m_{\varphi}v_{\rm vir}/\hbar$ is the wave vector where $v_{\rm vir}=10^{-3} c$, $\omega^2=m_{\varphi}c^4/\hbar^2+k_{\varphi}^2c^2$, the $\delta$ is the initial phase of the DM field and the effective wavelength is $\lambda_{\text{eff}}=\frac{\hbar}{m_{\text{eff}}c}$ with
\begin{align}
m^2_{\text{eff}}=m^2_{\varphi}+\frac{4\pi G\hbar^{2}}{c^4}\rho_{E}\tilde{\alpha}_{E}=m^2_{\varphi}+ (1.4\times 10^{-18} {\rm eV})^2\tilde{\alpha_{E}} \, .
\end{align}
One can see that the solution to the DM field is a sum of the background harmonic oscillation term and the local exponential
fluctuation term. The latter term comes from the effect of the Earth.

\subsection{The DM signal in single AI}

Here we consider the $\frac{\pi}{2}$-$\pi$-$\frac{\pi}{2}$ Raman AI \cite{PhysRevLett.67.181,RN264}. In the terrestrial AI experiments, the DM could change the atomic mass and the Earth's gravitational acceleration $g$. The DM also causes a modification of the electronic transition energy due to changes in the electronic mass $m_{e}$ and the electromagnetic fine structure constant $\alpha$. The modification of the atomic mass is
\begin{align}
m_{A}(\varphi)&=m_0(1+\alpha_{A}\varphi) \, ,\label{mchange}
\end{align}
where $\alpha_{A}$ is the DM charge for the atom as defined and calculated in \cite{RN18, RN110}. The  modification of the gravitational acceleration is
\begin{align}
g(\varphi)&= GM_E(\varphi)/r^2=g_{0}\big[1+\alpha_{E}\varphi+ \mathcal{O}(\varphi_0^2)+ \mathcal{O}(d_i^3)\big]\, ,\label{gchange}
\end{align}
The change in the electronic transition energy is
\begin{align}
\omega_{A}(\varphi)&\!=\omega_{A}\big[1+(d_{m_{e}}+\xi d_{e})\varphi\big] \, .\label{wchange}
\end{align}
For the $^{87}$Rb atom, $\xi\approx2.34$ is the relativistic correction factor \cite{RN50} and 
\begin{align} 
	\alpha_{87}&=9.1556685\times 10^{-1} d_{g}+8.3945\times 10^{-2}d_{\hat{m}}
	+2.54\times 10^{-4}d_{\delta m}\notag \\ 	&+2.339 \times 10^{-4}d_{m_{e}}	
+2.869\times 10^{-3} d_{e}\, . 
\end{align}
These modifications will be reflected in the AI's phase shift. The result for a single AI is calculated to be
	\begin{align}
		\phi&\!= \!-g_{0}T^2k_{\text{eff}}
		\!-\!k_{\text{eff}}\frac{c^2k_{\varphi}\alpha_{A}\varphi_{0}}{\omega^2}\bigg[\!\sin(k_{\varphi}r\!-\!2\omega T\!+\!\delta)\!-\!2\sin(k_{\varphi}r\!-\!\omega T\!+\!\delta)\!+\!\sin(k_{\varphi}r\!+\!\delta)\!\bigg]
		\notag \\
		&+\alpha_{A}\frac{2g_{0}k_{\rm eff}T}{\omega}\varphi_{0}\Big[\sin(k_{\varphi}r-\omega T+\delta)
		-\sin(k_{\varphi}r
		-2\omega T+\delta)\Big]
		\notag \\
		&\!+\!\left( \alpha_{E}\!+\!2\alpha_{A}\right) \frac{g_{0}k_{\text{eff}}}{\omega^{2}}\varphi_{0}\bigg[\cos(k_{\varphi}r+\delta)\!-\!2\cos(k_{\varphi}r-\omega T+\delta)\!+\!\cos(k_{\varphi}r-2\omega T+\delta)\bigg]
		\notag \\
		&-T^2k_{\text{eff}}\Bigg[\frac{\frac{7}{6}g_{0}T^2-(2v_{L}+v_{R})T}{\lambda_{\text{eff}}}+(1+\frac{r}{\lambda_{\text{eff}}})\Bigg]I(\frac{R_E}{\lambda_{\text{eff}}})g_{0}\alpha_{A}\alpha_{E}e^{-\frac{r}{\lambda_{\text{eff}}}}\, .
	\end{align}

\section{The ultralight scalar DM detection with ZAIGA \label{The ultralight scalar DM detection with ZAIGA}}

\subsection{The ZAIGA proposal}
\begin{figure}[htbp]
	\centering
	\includegraphics[width=0.98\textwidth]{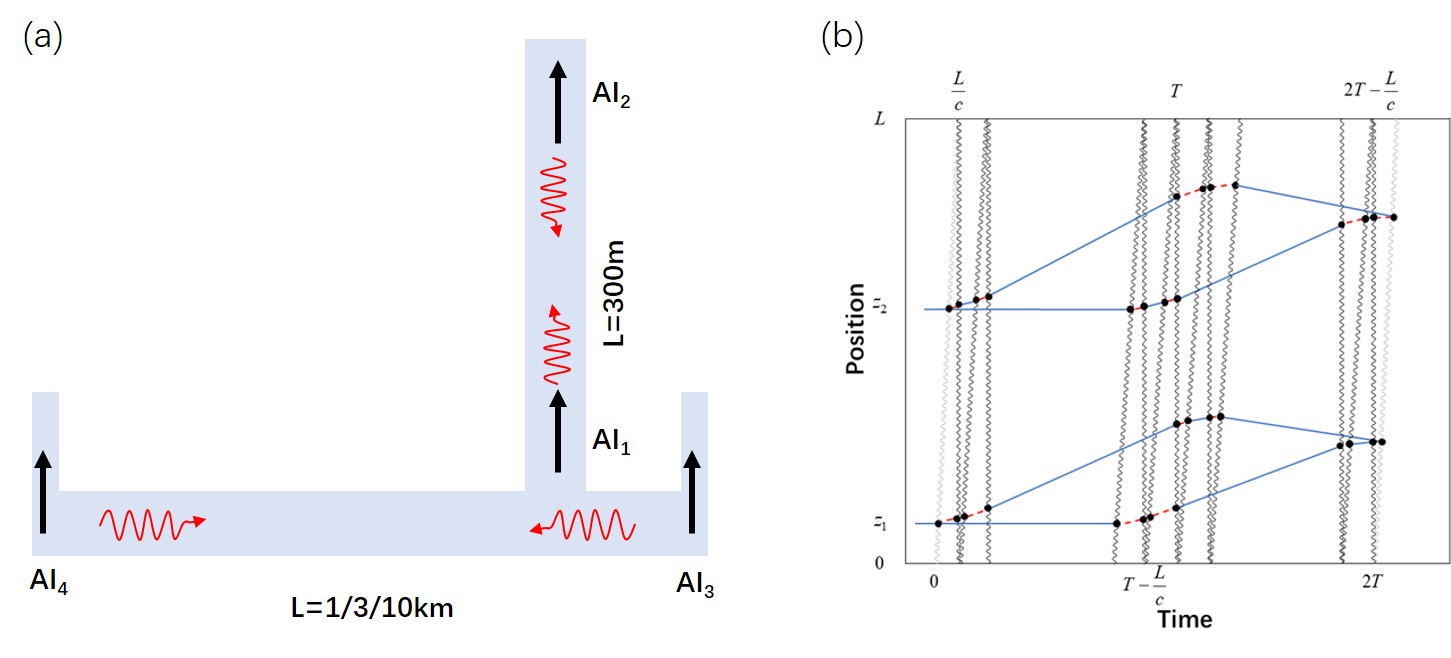}
	\caption{(a) the schematic diagram of ZAIGA. (b) the spacetime diagram of AIs. First the atomic beam is prepared in the ground state $|g\rangle$ and launched upwards with a velocity $v_{L}$. The $\frac{\pi}{2}$-pulse coherently splits the atom beam into a superposition of ground and excited states ($|g\rangle$ and $|e\rangle$) at the time $t=0$. The upper path will obtain a recoil momentum $\hbar k_{\text{eff}}$ (i.e., a recoil velocity $v_R$) compared with the lower path, where $\mathbf{k_{\text{eff}}}=\mathbf{k_{1}}-\mathbf{k_{2}}$. Then a $\pi$-pulse is used to transform the state $|g\rangle$ to
	$|e\rangle$ and the state $|e\rangle$ to $|g\rangle$ at the time $t=T$. Finally, a $\frac{\pi}{2}$-pulse is applied to recombine the two paths. The phase shift can be measured by detecting the number of atoms in ground or excited states.}    \label{apparatus}  
\end{figure}

  ZAIGA (Zhaoshan long-baseline atom interferometer gravitation antenna) is a proposed underground long-baseline atom interferometer facility \cite{RN106}. It includes a pair of 10-meter AIs vertically separated by 300 meters and a pair of 5-meter AIs horizontally separated by kilometers (See Fig. \ref{apparatus}). The spatially separated AIs are controlled by the same laser. An important advantage is that common-mode noises
  from laser phase fluctuations and platform vibrations can be cancelled out by taking differential phase measurements. The atomic shot noise could be suppressed by improving the atomic flux density. For the state-of-the-art technology of $ 10^{8}$ atoms/s, the phase sensitivity is $10^{-4}\text{rad}/\sqrt{\text{Hz}}$ \cite{RN284}. With the improvement of technology, the phase sensitivity could be improved by increasing flux density or adopting squeezed atomic states \cite{RN285}. 
  
  The facility sensitivity is also associated with the baseline length $L$ and the number of large momentum transfer (LMT) $n$. So it is possible to improve the sensitivity by adopting longer baseline and larger LMT. The LMT with 102 $\hbar k$ has been demonstrated and 1000 $\hbar k$ or larger is also possible  in the future  \cite{RN254,RN198}. The main technical parameters for ZAIGA are listed in Table \ref{table1} for the vertical configuration and Table \ref{table2} for the horizontal configuration. 

\begin{table}[bpth]
	\tbl{The technical parameters for a pair of vertically separated AIs}
	{\begin{tabular}{@{}cccccc@{}} \toprule
			&Free evolution&Phase sensitivity &Momentum&Integration time &Arm-
			\\
			&time (T)& & transfer (n) & ($t_{int}$) & length (L)
			\\ \colrule
			Near term  & 1.4s & \hphantom{0}$10^{-3}$rad/$\sqrt{\text{Hz}}$ & 4&$10^{4}$s&300m \\
			Future   & 1.4s & \hphantom{0}$10^{-4}$rad/$\sqrt{\text{Hz}}$ & $10^{4}$&$10^{6}$s&300m \\ \botrule
		\end{tabular} \label{table1}}
\end{table}
\begin{table}[bpth]
	\tbl{The technical parameters for a pair of horizontally separated AIs}
	{\begin{tabular}{@{}cccccc@{}} \toprule
			&Free evolution&Phase sensitivity &Momentum&Integration time &Arm-
			\\
			 &time (T)& & transfer (n) &($t_{int}$) & length (L)
			 \\ \colrule
			Near term  & 1s & \hphantom{0}$10^{-3}$rad/$\sqrt{\text{Hz}}$ & 4&$10^{4}$s&1km \\
			Future   & 1s & \hphantom{0}$10^{-7}$rad/$\sqrt{\text{Hz}}$ & $10^{3}$&$10^{6}$s&3km \\ \botrule
		\end{tabular} \label{table2}}
\end{table}

\subsection{The DM signal in two separated AIs}
The DM signal is measured by taking the differential phase between two separated AIs.

Firstly we consider the case of two vertically separated AIs. The phase shift of the first AI is given by 
\begin{align}
\phi_{1}&\!= \!-g_{1}T^2k_{\text{eff}}
\!-\!k_{\text{eff}}\frac{c^2k_{\varphi}\alpha_{A}\varphi_{0}}{\omega^2}\Big[\!\sin(k_{\varphi}R_{E}\!-\!2\omega T\!+\!\delta)\!-\!2\sin(k_{\varphi}R_{E}\!-\!\omega T\!+\!\delta)\!
\notag \\
&+\!\sin(k_{\varphi}R_{E}\!+\!\delta)\!\Big]
\!+\!\alpha_{A}\frac{2g_{1}k_{\rm eff}T}{\omega}\varphi_{0}\Big[\sin(k_{\varphi}R_{E}-\omega T+\delta)
-\sin(k_{\varphi}R_{E}
-2\omega T+\delta)\Big]
\notag \\
&\!+\!\left(\alpha_{E}\!+\!2\alpha_{A}\right)\! \frac{g_{1}k_{\text{eff}}}{\omega^{2}}\varphi_{0}\!\Big[\!\cos(k_{\varphi}R_{E}\!+\!\delta)\!\!-\!2\cos(k_{\varphi}R_{E}\!-\!\omega T\!+\!\delta)
\!\!+\!\!\cos(k_{\varphi}R_{E}\!-\!2\omega T\!+\!\delta)\!\Big]\!
\notag \\
&-T^2k_{\text{eff}}\bigg[\frac{\frac{7}{6}g_{1}T^2-(2v_{L}+v_{R})T}{\lambda_{\text{eff}}}+(1+\frac{R_{E}}{\lambda_{\text{eff}}})\bigg]I(\frac{R_{E}}{\lambda_{\text{eff}}})g_{1}\alpha_{A}\alpha_{E}e^{-\frac{R_{E}}{\lambda_{\text{eff}}}}\, ,
\label{phaseAI1}
\end{align} 
where $g_1$ denotes the gravitational acceleration at $z_1$.
The phase shift of the second AI located at $z_{2}$ is calculated to be
\begin{align}
\phi_{2}&\!= \!-g_{2}T^2k_{\text{eff}}
\!-\!k_{\text{eff}}\frac{c^2k_{\varphi}\alpha_{A}\varphi_{0}}{\omega^2}\Big[\!\sin(k_{\varphi}R_{E}\!-\!2\omega T\!+\!\delta)\!-\!2\sin(k_{\varphi}R_{E}\!-\!\omega T\!+\!\delta)
\notag \\
&\!+\!\sin(k_{\varphi}R_{E}\!+\!\delta)\!\Big]
+\alpha_{A}\frac{2g_{2}k_{\rm eff}T}{\omega}\varphi_{0}\Big[\sin(k_{\varphi}R_{E}-\omega T+\delta)
-\sin(k_{\varphi}R_{E}
-2\omega T+\delta)\Big]
\notag \\
&\!+\!\left(\alpha_{E}\!+\!2\alpha_{A}\right) \frac{g_{2}k_{\text{eff}}}{\omega^{2}}\varphi_{0}\!\Big[\!\cos(k_{\varphi}R_{E}\!+\!\delta)\!\!-\!\!2\cos(k_{\varphi}R_{E}\!-\!\omega T\!+\!\delta)\!\!+\!\!\cos(k_{\varphi}R_{E}\!-\!2\omega T\!+\!\delta)\!\Big]
\notag \\
&-T^2k_{\text{eff}}\bigg[\frac{\frac{7}{6}g_{1}T^2-(2v_{L}+v_{R})T}{\lambda_{\text{eff}}}+(1+\frac{R_{E}}{\lambda_{\text{eff}}})\bigg]I(\frac{R_{E}}{\lambda_{\text{eff}}})g_{2}\alpha_{A}\alpha_{E}e^{-\frac{R_{E}}{\lambda_{\text{eff}}}}
\notag \\
&\!+\!\frac{(d_{m_{e}}\!\!+\!\xi d_{e})\omega_{A}}{\omega}\varphi_{0}\bigg[\sin\big[k_{\varphi}R_{E}\!-\!\omega \big(T\!+\!\frac{L}{c}\big)\!+\!\delta\big]\!-\!\sin\big[k_{\varphi}R_{E}\!-\!\omega \big(T\!-\!(n\!-\!1)\frac{L}{c}\big)\!+\!\delta\big]
\notag \\
&-\sin\big[k_{\varphi}R_{E}-\frac{\omega n L}{c}+\delta\big] +\sin\big[k_{\varphi}R_{E}+\delta\big]-\sin\big[k_{\varphi}R_{E}-\omega \big(2T+\frac{L}{c}\big)+\delta\big]
\notag \\
&+\sin\big[k_{\varphi}R_{E}-\omega \big(2T-(n-1)\frac{L}{c}\big)+\delta\big]
+\sin\big[k_{\varphi}R_{E}-\omega \big(T+\frac{nL}{c}\big)+\delta\big]
\notag \\
&-\sin\big[k_{\varphi}R_{E}-\omega T+\delta\big]\bigg] \, ,
\label{phaseAI2}
\end{align}
where $g_2$ is the gravitational acceleration at $z_2$.

So the differential phase $\Phi^{\rm v}_{\text{sum}}$ for two vertically separated AIs is $\phi_{2}-\phi_{1}$ , which can be rewritten into the following form  
\begin{align}
\Phi^{\rm v}_{\text{sum}}& =-(g_2-g_{1})T^2k_{\text{eff}}+\Phi_{\rm exp}+\Phi_{\rm osc}+\Phi_{\rm eng} \, ,
\label{vsignal}
\end{align}
with
\begin{align}
	\Phi_{\rm exp}&= -T^2k_{\text{eff}}\Bigg[\frac{\frac{7}{6}g_{1}T^2-(2v_{L}+v_{R})T}{\lambda_{\text{eff}}}+(1+\frac{R_{E}}{\lambda_{\text{eff}}})\Bigg]I(\frac{R_{E}}{\lambda_{\text{eff}}})(g_{2}-g_{1})\alpha_{A}\alpha_{E}e^{-\frac{R_{E}}{\lambda_{\text{eff}}}} \, , \label{static term}
\end{align}
\begin{align}
	\Phi_{\rm osc}& = \alpha_{A}\frac{2(g_{2}-g_{1})k_{\rm eff}T}{\omega}\varphi_{0}\Big[\sin(k_{\varphi}R_{E}-\omega T+\delta)
	-\sin(k_{\varphi}R_{E}\!
	-\!2\omega T\!+\!\delta)\Big]
	\notag \\
	&+ \big(\alpha_{E}+2\alpha_{A}\big) \frac{(g_{2}-g_{1})k_{\text{eff}}}{\omega^{2}}\varphi_{0}\Big[\cos(k_{\varphi}R_{E}+\delta)-2\cos(k_{\varphi}R_{E}-\omega T+\delta)
	\notag \\
	&+\cos(k_{\varphi}R_{E}-2\omega T+\delta)\Big]\, ,
\end{align}
and
\begin{align}
	\Phi_{\rm eng}&= 
	\frac{(d_{m_{e}}\!\!+\!\xi d_{e})\omega_{A}}{\omega}\!\bigg[\!\sin\!\big[k_{\varphi}R_{E}\!-\!\omega \big(T\!+\!\frac{L}{c}\big)\!+\!\delta\big]\!-\!\sin\big[k_{\varphi}R_{E}\!-\!\omega \big(T\!-\!(n\!-\!1)\frac{L}{c}\big)\!+\!\delta\big]
	\notag \\
	&-\sin\big[k_{\varphi}R_{E}-\frac{\omega n L}{c}+\delta\big] +\sin\big[k_{\varphi}R_{E}+\delta\big]-\sin\big[k_{\varphi}R_{E}-\omega \big(2T+\frac{L}{c}\big)+\delta\big]
	\notag \\
	&+\sin\big[k_{\varphi}R_{E}-\omega \big(2T-(n-1)\frac{L}{c}\big)+\delta\big]
	+\sin\big[k_{\varphi}R_{E}-\omega \big(T+\frac{nL}{c}\big)+\delta\big]
	\notag \\
	&-\sin\big[k_{\varphi}R_{E}-\omega T+\delta\big]\bigg]\, .  \label{transitional phase}
\end{align}

According to Eqs. (\ref{mchange}) and (\ref{gchange}), the atomic mass and the gravitational acceleration are changed by the DM. $\Phi_{\rm osc}$ stands for the contribution due to changes in the atomic mass and the gravitational acceleration caused by the oscillatory term of the DM field. $\Phi_{\rm exp}$ denotes the contribution due to changes in the atomic mass and the gravitational acceleration caused by the exponential fluctuation term of the DM field. They are both proportional to $g_2-g_1$, and are functions of all the five DM coupling parameters. Lastly, according to Eq. (\ref{wchange}), the electronic transition energy is also influenced by the DM. Then, we use $\Phi_{\rm eng}$ to denote the corresponding contribution to the differential phase, which is surely independent of the gravitational acceleration, and is a function of $d_e$ and $d_{m_{e}}$. It has already been discussed in papers \cite{RN110,RN20}.

Secondly, we consider the case of two horizontally separated AIs. To be simple, we ignore the horizontal gradient in the Earth's gravity field. Doing the similar calculation of phase shifts for the two AIs as before, one can find that the phase shifts are also given by Eqs. (\ref{phaseAI1}) and (\ref{phaseAI2}) except $g_1$ = $g_2$.  Thus, the differential phase for two horizontally separated AIs is found to be  
 \begin{align}
  \Phi^{\rm h}_{\text{sum}}&=\phi_{2}-\phi_{1}=\Phi_{\rm eng}\,.
 \end{align}
In other words, the DM signal comes only from the modification of the electronic transitional energy. 

 \begin{figure}[htbp]
	\centering
	\includegraphics[width=0.98\textwidth]{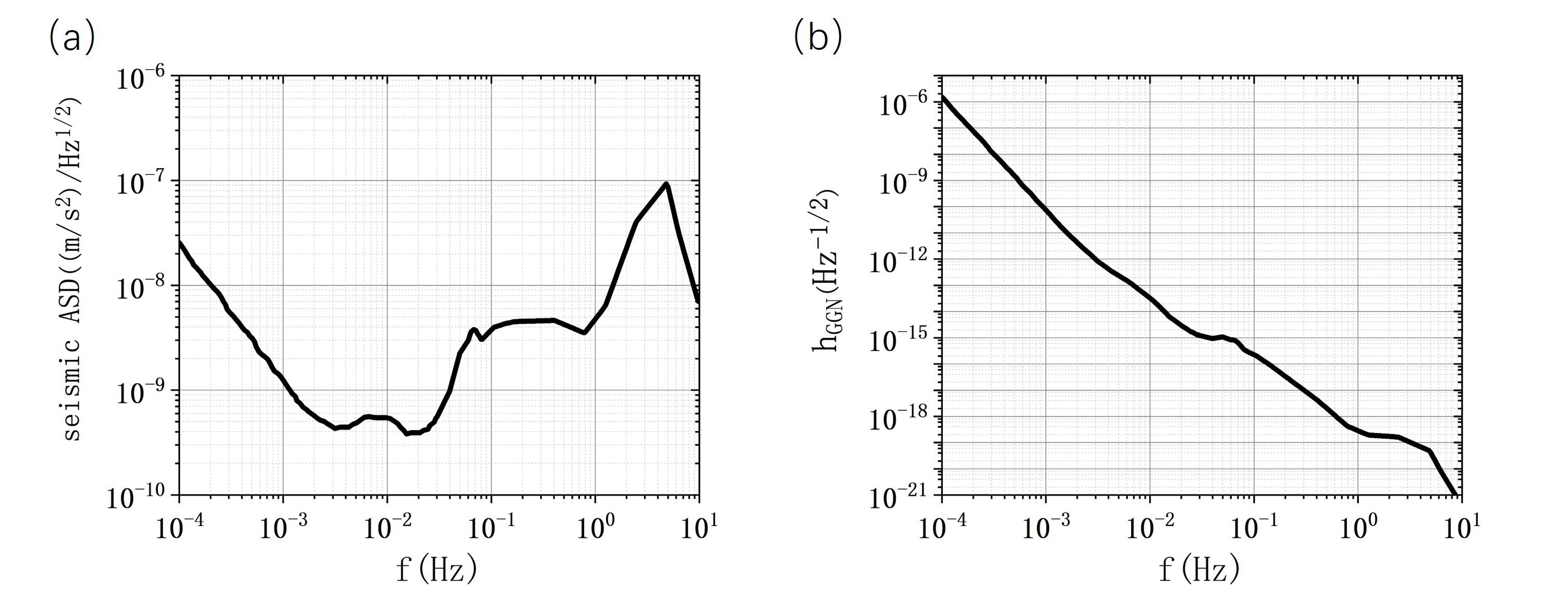}
	\caption{(a) Seismic acceleration amplitude spectral density given by the NLNM \cite{Peterson93,Zurn07}. (b) The corresponding strain amplitude spectral density.} \label{GGN}
\end{figure}
\begin{figure}[htbp]
	\centering
	\includegraphics[width=0.49\textwidth]{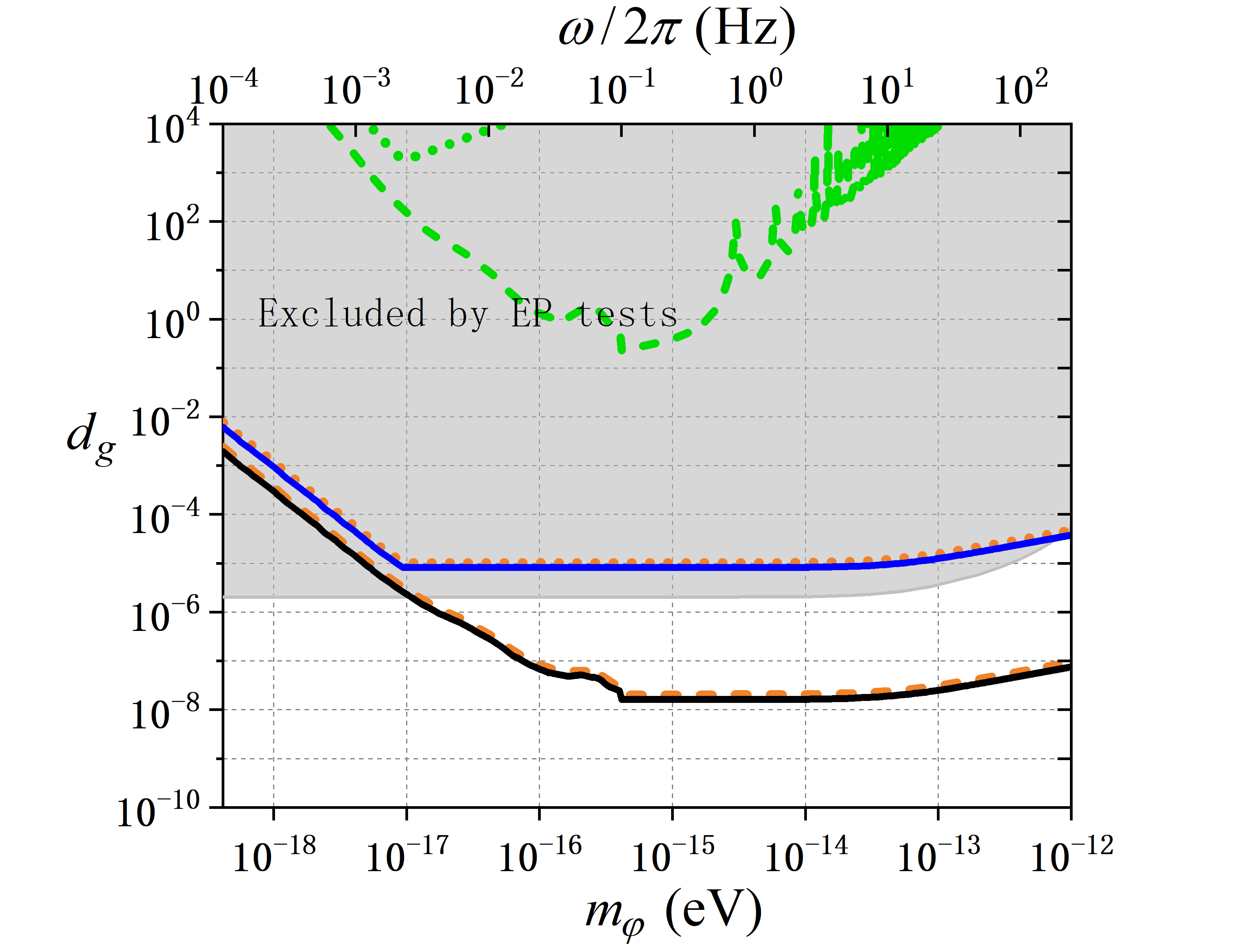}
	\vspace{0.5cm}
	\includegraphics[width=0.49\textwidth]{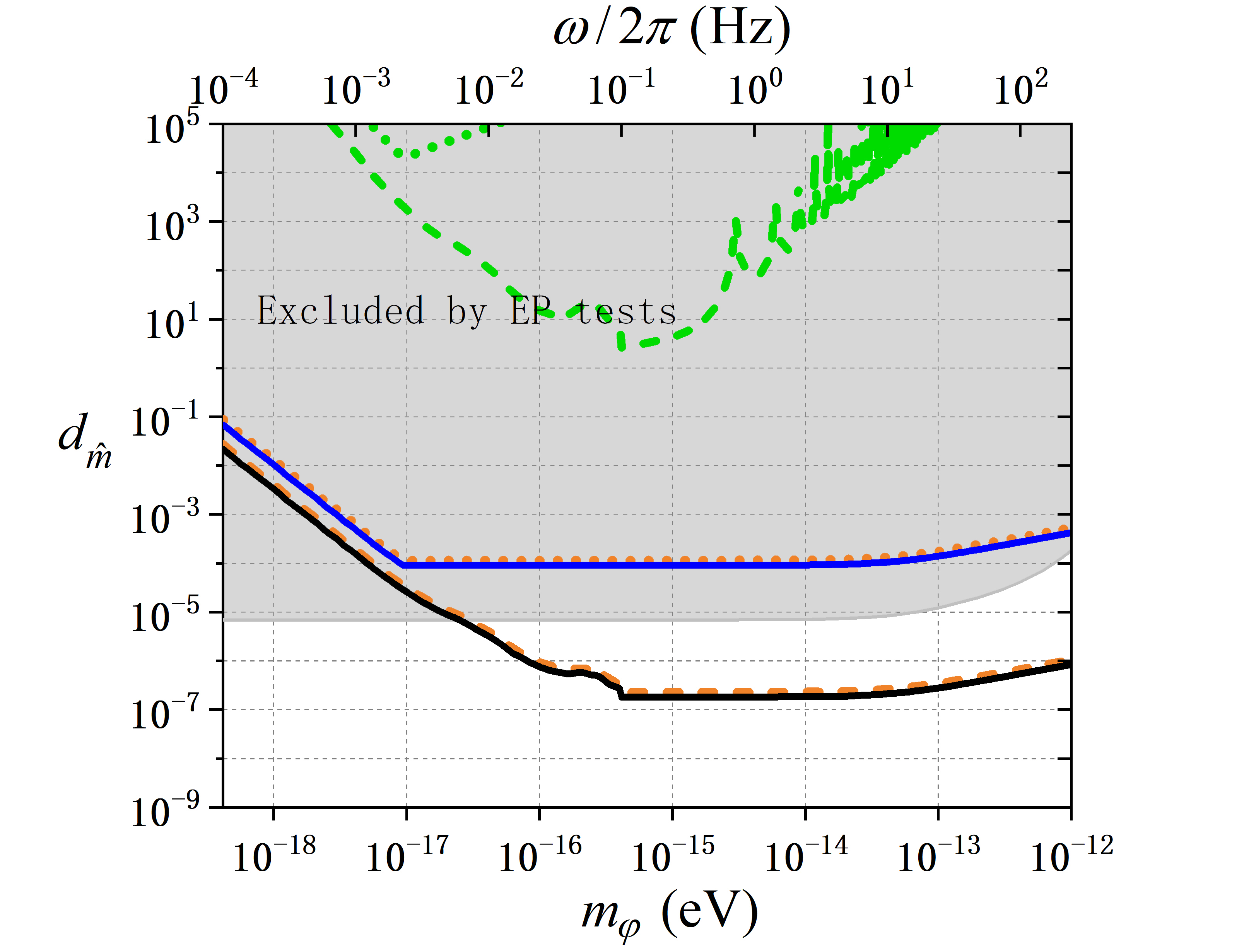}
	\includegraphics[width=0.49\textwidth]{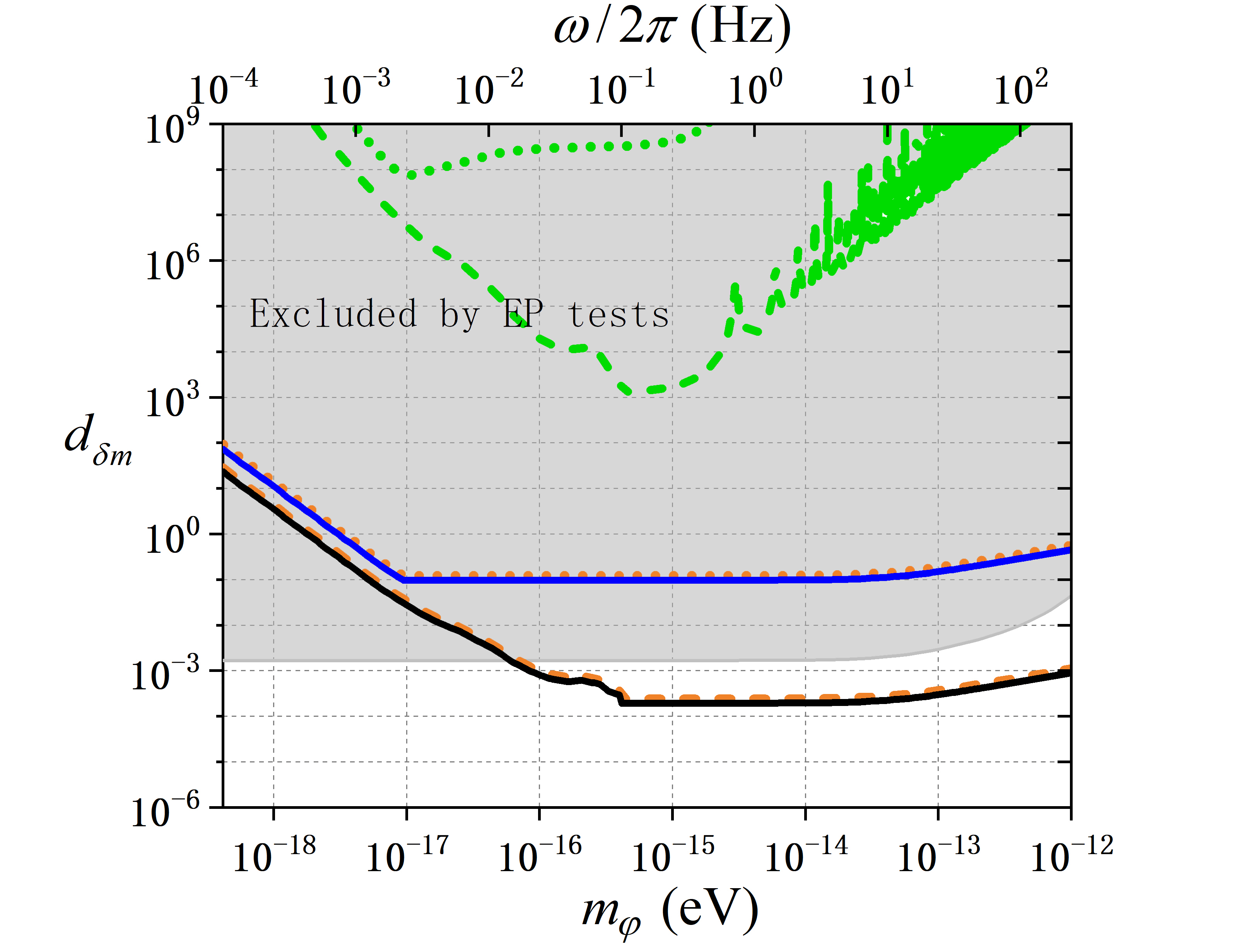}
	\vspace{0.5cm}
	\includegraphics[width=0.49\textwidth]{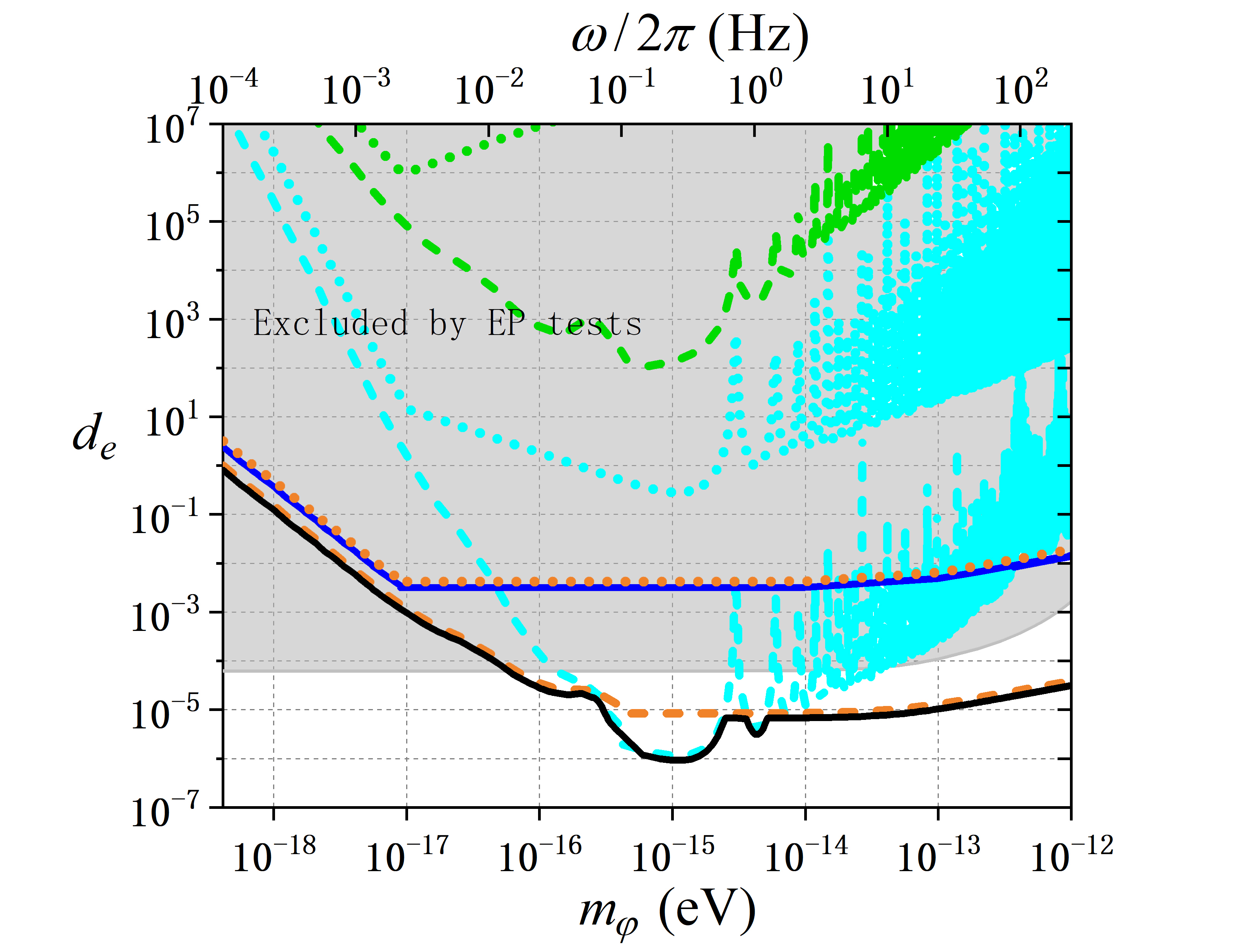} 
	\includegraphics[width=0.49\textwidth]{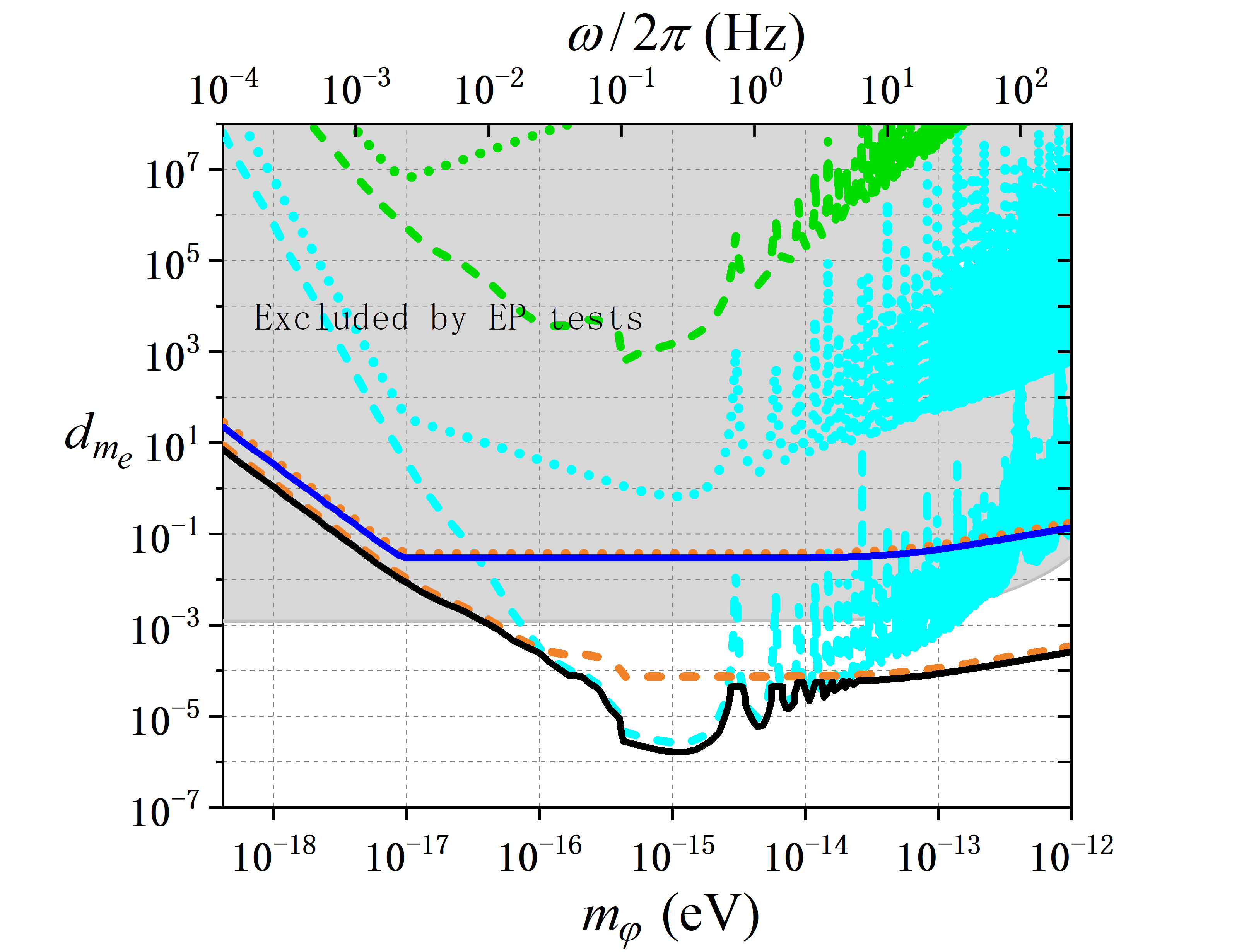}
	\caption{Constraints on DM coupling parameters by a pair of vertically separated AIs, assuming a 50-time mitigation in the NN. The blue and black solid lines are the overall constraints set by the near-term ZAIGA and the future ZAIGA. The gray areas denote the parameter regions excluded by the MICROSCOPE space mission\cite{Touboul17:MICROSCOPE,Berge18MICROSCOPEConstraints}. Dotted lines and dashed lines denote the various contributions to the corresponding overall constraints for the near-term ZAIGA and the future ZAIGA, respectively. The green color is for $\bar{\Phi}_{\rm osc}$, the cyan color is for $\bar{\Phi}_{\rm eng}$, and the orange color is for $\Phi_{\rm exp}$. \label{constraint by vertical direction}}
\end{figure}
\begin{figure}[htbp]
	\centering
	\includegraphics[width=0.49\textwidth]{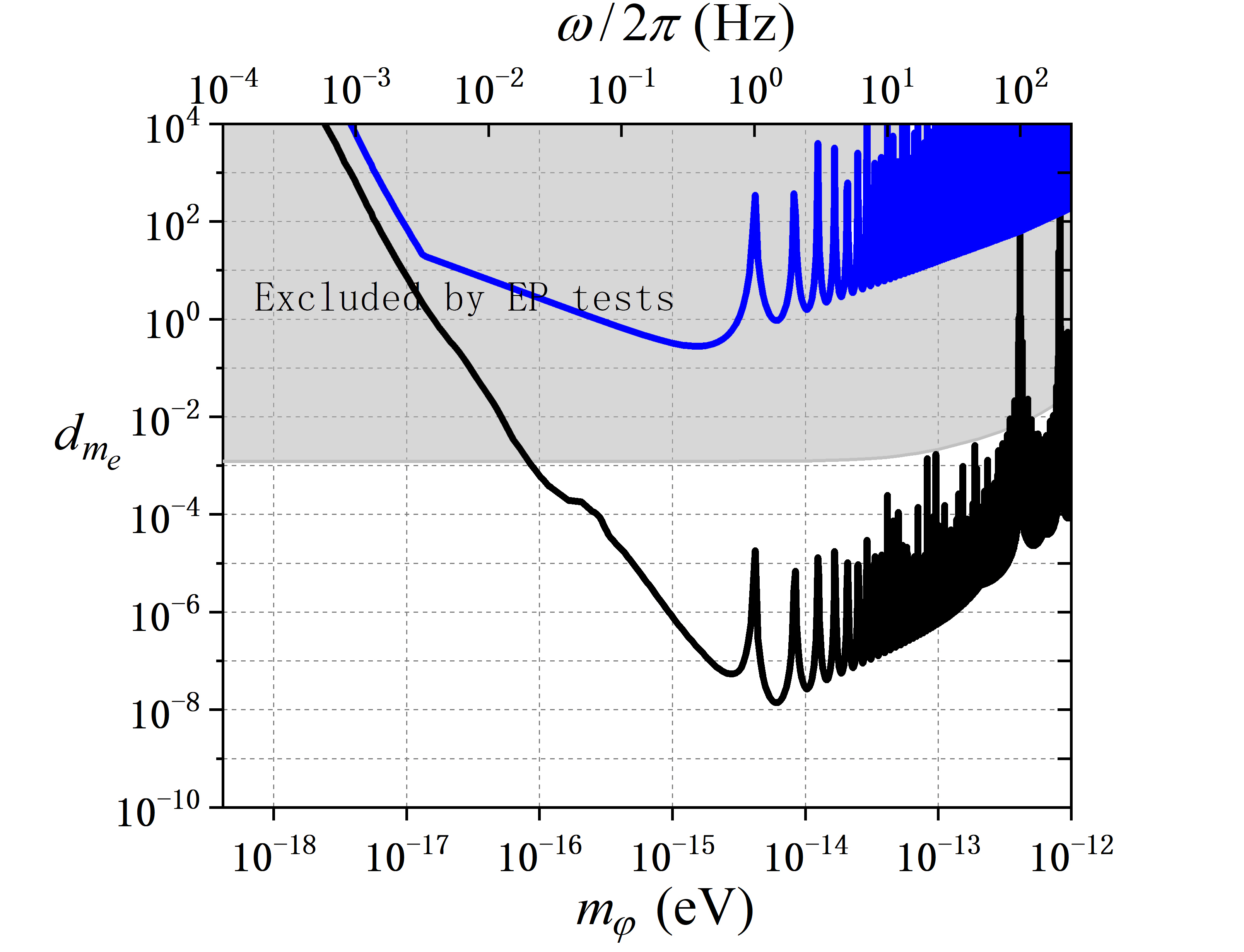}
	\includegraphics[width=0.49\textwidth]{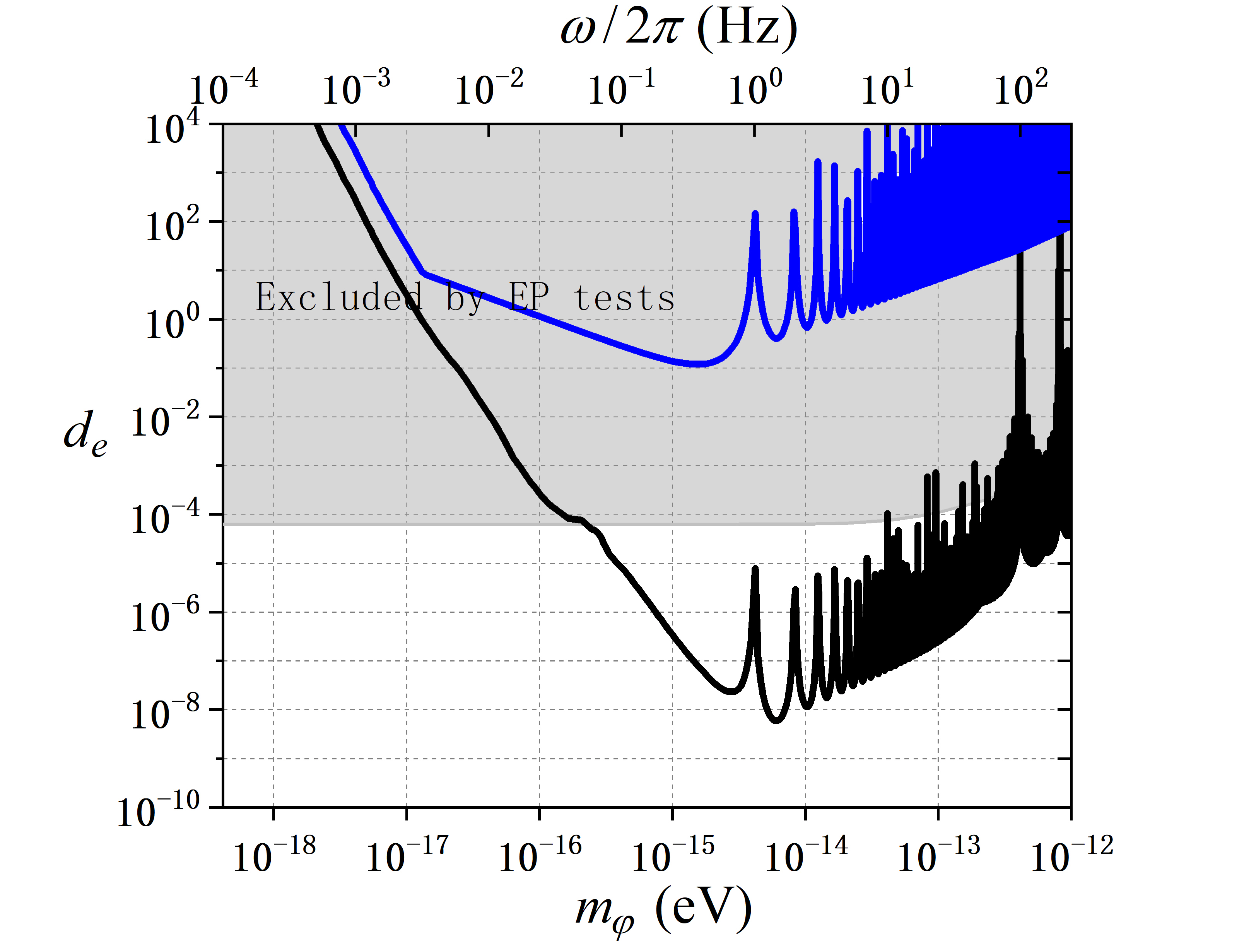}
	\caption{Constraints on DM coupling parameters by a pair of horizontally separated AIs, assuming a 50-time mitigation in NN. The blue and black solid lines are the overall constraints set by the near-term ZAIGA and the future ZAIGA. } \label{horizontal constraint}
\end{figure}

\section{Constraints on the DM coupling parameters \label{Constraints on the coupling parameters}}

We constrain only one parameter every time by setting the rest four parameters to zero. This method has been used in many papers \cite{RN2,Berge18MICROSCOPEConstraints,RN37}. 

Let us first discuss the vertical configuration of ZAIGA, where a pair of 10-meter AIs are vertically separated by 300 meters. The DM signal is given by Eq. (\ref{vsignal}), which is a sum of $\Phi_{\rm exp}$, $\Phi_{\rm osc}$, and $\Phi_{\rm eng}$. Since  $\Phi_{\rm osc}$ and $\Phi_{\rm eng}$ are oscillatory in the initial phase $\delta$, it is helpful to use the corresponding signal amplitudes defined by $\bar{\Phi}_{\rm osc}=\big(2\int_{0}^{2\pi}\Phi^{2}_{\rm osc}/2\pi d\delta\big)^{1/2}$, and  $\bar{\Phi}_{\rm eng}=\big(2\int_{0}^{2\pi}\Phi^{2}_{\rm eng}/2\pi d\delta\big)^{1/2}$. As discussed in the paper \cite{RN106}, the sensitivity of ZAIGA is determined by the atomic shot noise and the Newtonian gravity gradient noise (NN). The atomic shot noise is determined by the technical parameters in Table \ref{table1}. The NN is a crucial background noise for any terrestrial gravitational wave detector, which varies widely for different experimental sites. Since we have not started any real-time NN measurement, here we use the well-known NLNM (the New Low Noise Model) \cite{Peterson93,Zurn07} to estimate the NN, as depicted in Fig. \ref{GGN}. On the other hand, according to the paper \cite{Harms19}, it is possible that the NN could be mitigated by 50 times by using dedicated arrays of auxiliary sensors. Then, constraints on $d_{g}$, $d_{\hat{m}}$, $d_{\delta m}$, $d_{e}$ and $d_{m_{e}}$ are depicted in Fig. \ref{constraint by vertical direction}, assuming a 50-time mitigation in the NN. It is clear that curves for the near-term ZAIGA are determined by the NN below $2\times 10^{-3}$Hz and by the shot noise above that frequency. Curves for the future ZAIGA are determined by the NN below 0.1Hz and by the shot noise above that frequency. The blue and black solid lines are the overall constraints set by the near-term ZAIGA and the future ZAIGA. The gray areas denote the parameter regions excluded by the MICROSCOPE space mission \cite{Touboul17:MICROSCOPE,Berge18MICROSCOPEConstraints}. 

In Fig. \ref{constraint by vertical direction}, in order to obtain the overall constraints in the full DM mass range, we need to first find out the contributions to the overall constraint from $\bar{\Phi}_{\rm eng}$, $\bar{\Phi}_{\rm osc}$ and $\Phi_{\rm exp}$ individually. The green lines are the constraints set by $\bar{\Phi}_{\rm osc}$, and the orange lines are the constraints set by $\Phi_{\rm exp}$ in all the five pictures. In the pictures for $d_{e}$ and $d_{m_{e}}$,  the cyan lines denote the constraints set by $\bar{\Phi}_{\rm eng}$. Then, by connecting the dominant constraint curves in different DM mass range, one can get the overall constraint curves, which are denoted by blue and black solid curves. For the near-term ZAIGA, constraints set by $\Phi_{\rm exp}$ are always the dominant ones. For the future ZAIGA, constraints set by $\Phi_{\rm exp}$ are the dominant ones over most DM mass ranges. Only in the pictures for $d_{e}$ and $d_{m_{e}}$, in the DM mass range $10^{-16} {\rm eV}-10^{-14} {\rm eV}$, the dominant constraints are set by $\bar{\Phi}_{\rm eng}$. In one word, the final constraints for all five DM coupling parameters set by the near-term ZAIGA are worse than the MICROSCOPE's constraints by one to two orders of magnitude. However, for the future ZAIGA, the final constraints are better than the MICROSCOPE's constraints by one to two orders of magnitude above the $10^{-16} {\rm eV}$ DM mass range.

Next, let us consider the horizontal configuration of ZAIGA, where a pair of 5-meter AIs are  horizontally separated by 1000 meters. The DM signal is given by $\bar{\Phi}_{\rm eng}$. Since $\bar{\Phi}_{\rm eng}$ depends on $d_e$ and $d_{m_{e}}$, we could only constrain $d_{e}$ and $d_{m_{e}}$, which are shown in Fig. \ref{horizontal constraint}, again assuming a 50-time mitigation in the NN. For the near-term ZAIGA, curves are determined by the NN below $3\times 10^{-3}$Hz and by the shot noise above that frequency. Curves for the future ZAIGA are determined by the NN below 1.0Hz and by the shot noise above that frequency. It is clear that the final constraints from the near-term ZAIGA are worse than the MICROSCOPE's constraints. On the other hand, for the future ZAIGA, the final constraints are better than the MICROSCOPE's constraints above the $10^{-16} {\rm eV}$ DM mass range. Especially, in the DM mass range $10^{-15} {\rm eV}-10^{-13} {\rm eV}$, the constraints can be improved by more than three orders of magnitude.

\section{Conclusion and discussion\label{Conclusion and discussion}}

In this paper, we discuss the ultralight scalar DM detection with ZAIGA. According to a popular scalar DM model, the DM field couples to the standard model matter through five coupling parameters. We calculate the DM signals in ZAIGA, and give the expected constraints for the five DM coupling parameters. It turns out that our proposed constraints could be several orders of magnitude better than the ones set by the MICROSCOPE space mission. For the vertical configuration of ZAIGA, the advantage is that all the five DM coupling parameters can be constrained, and the disadvantage is that the arm-length is not easy to extend. For the horizontal configuration of ZAIGA, the advantage is that the arm-length is relatively easy to extend, and the disadvantage is that only $d_{e}$ and $d_{m_{e}}$ can be constrained. In summary, ZAIGA shows an impressive potential on the ultralight scalar DM detection over a wide mass range, and can join with other long-baseline atomic sensor schemes to form a network of the ultralight scalar DM detection.

\section*{Acknowledgments}

 This work was supported by the National Key Research and Development Program of China under Grant No. 2016YFA0302002, and the Strategic Priority Research Program of the Chinese Academy of Sciences under Grant No. XDB21010100.

\bibliographystyle{ws-ijmpd}
\bibliography{sample}

\begin{thebibliography}{10}

\bibitem{1}
G.~Bertone and D.~Hooper, {\em Rev. Mod. Phys.} {\bf 90}  (2018)   045002.

\bibitem{2}
S.~W. {Allen}, R.~W. {Schmidt}, A.~C. {Fabian} and H.~{Ebeling}, {\em Mon. Not.
  R. Astron. Soc.} {\bf 342}  (2003) 287.

\bibitem{3}
C.~Mondino, A.-M. Taki, K.~Van~Tilburg and N.~Weiner, {\em Phys. Rev. Lett.}
  {\bf 125}  (2020)   111101.

\bibitem{cosmological}
M.~S. Safronova, D.~Budker, D.~DeMille, D.~F.~J. Kimball, A.~Derevianko and
  C.~W. Clark, {\em Rev. Mod. Phys.} {\bf 90}  (2018)   025008.

\bibitem{RN405}
R.~Agnese~{\it et al.}, {\em Phys. Rev. Lett.} {\bf 121}  (2018)   051301.

\bibitem{RN403}
E.~Aprile~{\it et al.}, {\em Phys. Rev. Lett.} {\bf 121}  (2018)   111302.

\bibitem{RN404}
D.~S. Akerib~{\it et al.}, {\em Phys. Rev. Lett.} {\bf 122}  (2019)   131301.

\bibitem{RN17}
A.~A. Geraci and A.~Derevianko, {\em Phys. Rev. Lett.} {\bf 117}  (2016)
  261301.

\bibitem{Derevianko14}
A.~Derevianko and M.~Pospelov, {\em Nat. Phys.} {\bf 10}  (2014)   933.

\bibitem{RN2}
A.~Arvanitaki, J.~Huang and K.~Van~Tilburg, {\em Phys. Rev. D} {\bf 91}  (2015)
    015015.

\bibitem{Roberts20}
B.~M. Roberts {\em et~al.}, {\em New J. Phys.} {\bf 22}  (2020)   093010.

\bibitem{Beloy21Nature:ACDM}
K.~Beloy {\em et~al.}, {\em Nature} {\bf 591}  (2021)   564.

\bibitem{RN45}
P.~W. Graham, D.~E. Kaplan, J.~Mardon, S.~Rajendran and W.~A. Terrano, {\em
  Phys. Rev. D} {\bf 93}  (2016)   075029.

\bibitem{Manley21PRL:OptomechanicalAccelerometerVectorDM}
J.~Manley {\em et~al.}, {\em Phys. Rev. Lett.} {\bf 126}  (2021)   061301.

\bibitem{Carney21}
D.~Carney {\em et~al.}, {\em New J. Phys.} {\bf 23}  (2021)   023041.

\bibitem{Obata18}
I.~Obata, T.~Fujita and Y.~Michimura, {\em Phys. Rev. Lett.} {\bf 121}  (2018)
   161301.

\bibitem{RN129}
A.~A. Geraci, C.~Bradley, D.~F. Gao, J.~Weinstein and A.~Derevianko, {\em Phys.
  Rev. Lett.} {\bf 123}  (2019)   031304.

\bibitem{Savalle21}
E.~Savalle {\em et~al.}, {\em Phys. Rev. Lett.} {\bf 126}  (2021)   051301.

\bibitem{Kennedy20}
C.~J. Kennedy {\em et~al.}, {\em Phys. Rev. Lett.} {\bf 125}  (2020)   201302.

\bibitem{Stadnik15:LaserInterferometry}
Y.~V. Stadnik and V.~V. Flambaum, {\em Phys. Rev. Lett.} {\bf 114}  (2015)
  161301.

\bibitem{Stadnik16}
Y.~V. Stadnik and V.~V. Flambaum, {\em Phys. Rev. A} {\bf 93}  (2016)   063630.

\bibitem{Grote19PRR:GWDM}
H.~Grote and Y.~V. Stadnik, {\em Phys. Rev. Research} {\bf 1}  (2019)   033187.

\bibitem{Pierce18PRLDPGW}
A.~Pierce, K.~Riles and Y.~Zhao, {\em Phys. Rev. Lett.} {\bf 121}  (2018)
  061102.

\bibitem{Nagano19}
K.~Nagano, T.~Fujita, Y.~Michimura and I.~Obata, {\em Phys. Rev. Lett.} {\bf
  123}  (2019)   111301.

\bibitem{Guo19:DPLIGO}
H.-K. Guo, K.~Riles, F.-W. Yang and Y.~Zhao, {\em Commun. Phys.} {\bf 2}
  (2019)   155.

\bibitem{Morisaki19}
S.~Morisaki and T.~Suyama, {\em Phys. Rev. D} {\bf 100}  (2019)   123512.

\bibitem{Morisaki21PRD:GWvectorDM}
S.~Morisaki {\em et~al.}, {\em Phys. Rev. D} {\bf 103}  (2021)   L051702.

\bibitem{Michimura20PRD:GWvectorDM}
Y.~Michimura {\em et~al.}, {\em Phys. Rev. D} {\bf 102}  (2020)   102001.

\bibitem{RN78}
A.~D. Cronin, J.~Schmiedmayer and D.~E. Pritchard, {\em Rev. Mod. Phys.} {\bf
  81}  (2009) 1051.

\bibitem{RN190}
P.~Asenbaum, C.~Overstreet, M.~Kim, J.~Curti and M.~A. Kasevich, {\em Phys.
  Rev. Lett.} {\bf 125}  (2020)   191101.

\bibitem{RN194}
L.~Morel, Z.~Yao, P.~Cladé and S.~Guellati-Khélifa, {\em Nature} {\bf 588}
  (2020) 61.

\bibitem{RN286}
G.~M. Tino, {\em Quantum Sci. Technol.} {\bf 6}  (2021)   024014.

\bibitem{RN255}
L.~Badurina~{\it et al.}, {\em J. Cosmol. Astropart. Phys.} {\bf 5}  (2020)
  011.

\bibitem{RN247}
M.~Abe~{\it et al.}, {\em Quantum Sci. Technol.} {\bf 6}  (2021)   044003.

\bibitem{RN187}
B.~Canuel~{\it et al.}, {\em Sci. Rep.} {\bf 8}  (2018)   14064.

\bibitem{RN256}
B.~Canuel~{\it et al.}, {\em Class. Quantum Grav.} {\bf 37}  (2020)   225017.

\bibitem{RN106}
M.-S. Zhan~{\it et al.}, {\em Int. J. Mod. Phys. D} {\bf 29}  (2020)   1940005.

\bibitem{RN18}
T.~Damour and J.~F. Donoghue, {\em Phys. Rev. D} {\bf 82}  (2010)   084033.

\bibitem{RN16}
A.~Hees, O.~Minazzoli, E.~Savalle, Y.~V. Stadnik and P.~Wolf, {\em Phys. Rev.
  D} {\bf 98}  (2018)   064051.

\bibitem{RN110}
W.~Zhao, D.~F. Gao, J.~Wang and M.~S. Zhan, {\em arXiv:2102.02391}   (2021).

\bibitem{PhysRevLett.67.181}
M.~Kasevich and S.~Chu, {\em Phys. Rev. Lett.} {\bf 67}  (1991) 181.

\bibitem{RN264}
M.~Kasevich and S.~Chu, {\em Appl. Phys. B} {\bf 54}  (1992) 321.

\bibitem{RN50}
V.~V. Flambaum and A.~F. Tedesco, {\em Phys. Rev. C} {\bf 73}  (2006)   055501.

\bibitem{RN284}
P.~Treutlein, K.~Y. Chung and S.~Chu, {\em Phys. Rev. A} {\bf 63}  (2001)
  053821.

\bibitem{RN285}
O.~Hosten, N.~J. Engelsen, R.~Krishnakumar and M.~A. Kasevich, {\em Nature}
  {\bf 529}  (2016) 505.

\bibitem{RN254}
H.~M{\"u}ller, S.~W. Chiow, Q.~Long, S.~Herrmann and S.~Chu, {\em Phys. Rev.
  Lett.} {\bf 100}  (2008)   180405.

\bibitem{RN198}
S.~W. Chiow, T.~Kovachy, H.~C. Chien and M.~A. Kasevich, {\em Phys. Rev. Lett.}
  {\bf 107}  (2011)   130403.

\bibitem{RN20}
A.~Arvanitaki, P.~W. Graham, J.~M. Hogan, S.~Rajendran and K.~Van~Tilburg, {\em
  Phys. Rev. D} {\bf 97}  (2018)   075020.

\bibitem{Peterson93}
J.~R. Peterson, {\em Observations and modeling of seismic background noise},
  U.S. Geol. Surv. Open-File Report 93-322  (1993).

\bibitem{Zurn07}
W.~Zürn and E.~Wielandt, {\em Geophys. J. Int.} {\bf 168}  (2007)   647.

\bibitem{Touboul17:MICROSCOPE}
P.~Touboul {\em et~al.}, {\em Phys. Rev. Lett.} {\bf 119}  (2017)   231101.

\bibitem{Berge18MICROSCOPEConstraints}
J.~Bergé {\em et~al.}, {\em Phys. Rev. Lett.} {\bf 120}  (2018)   141101.

\bibitem{RN37}
N.~Leefer, A.~Gerhardus, D.~Budker, V.~V. Flambaum and Y.~V. Stadnik, {\em
  Phys. Rev. Lett.} {\bf 117}  (2016)   271601.

\bibitem{Harms19}
J.~Harms, {\em Living Rev. Relativ.} {\bf 22}  (2019)  ~6.

\end{thebibliography}
\end{document}